\documentclass[amssymb,12pt,longbibliography,nofootinbib]{revtex4-1}
\usepackage{graphicx}
\usepackage{appendix} 
\usepackage{hyperref}
\usepackage{bm,amsmath}
\usepackage{amsthm}
\usepackage{setspace}
\usepackage{slashed,mathrsfs}
\usepackage{xcolor}

\def\BLO{ B ({\cal H}, {\mathbb C}) }

\def\DIM{ N }

\def\HER{ {\rm Her} ({\cal H}, {\mathbb C}) }

\def\PSD{ {\rm Her}^{\ge 0} ({\cal H}, {\mathbb C}) }

\begin{document}
\rightline{\href{https://iopscience.iop.org/article/10.1088/1572-9494/acf304}{\tt Commun.~Theor.~Phys.~{\bf 75} 105102 (2023)}}
\vspace{0.2in}
\title{Nonlinear and non-CP gates for Bloch vector amplification}
\author{\textsc{Michael R. Geller}}
\affiliation{Center for Simulational Physics, University of Georgia, Athens, Georgia 30602, USA}
\date{September 22, 2023}

\begin{abstract}
\vskip 0.5in
\centerline{\bf Abstract}
\vskip 0.1in
\begin{spacing}{0.9}
Any state ${\bf r} = (x,y,z)$ of a qubit, written in the Pauli basis and initialized in the pure state ${\bf r} = (0,0,1)$, can be prepared by composing three quantum operations: two unitary rotation gates to reach a pure state ${\bf r} = (x^2 + y^2 + z^2)^{-\frac{1}{2}} \times (x,y,z)$ on the Bloch sphere, followed by a depolarization gate to decrease $| {\bf r} |$. Here we discuss the complementary state-preparation protocol for qubits initialized at the center of the Bloch ball, ${\bf r}=0$, based on increasing or amplifying $| {\bf r} |$ to its desired value, then rotating. Bloch vector amplification increases purity and decreases entropy. Amplification can be achieved with a linear Markovian CPTP channel by placing the channel's fixed point away from ${\bf r}=0$, making it nonunital, but the resulting gate suffers from a critical slowing down as that fixed point is approached. Here we consider alternative designs based on linear and nonlinear Markovian PTP channels, which offer benefits relative to linear CPTP channels, namely fast Bloch vector amplification without deceleration. These operations simulate a reversal of the thermodynamic arrow of time for the qubit and would provide striking experimental demonstrations of non-CP dynamics.
\end{spacing}
\end{abstract}
\maketitle

Several papers have explored the use of real or effective quantum nonlinearity for information processing \cite{MielnikJMP80,PhysRevLett.81.3992,BechmannPLA98,9802051,9803019,9811036,0309189,0502072,BrunPRL09,BennettPRL09,13033537,13030371,13107301,150706334,200907800,220613362,XuPRR22}. Nonlinear master equations have also been  frequently discussed in open systems theory 
\cite{GisinJPA81,AlickiJSP83,Breuer2002,190810145, GisinJPA92,BrodyPRL12,14057165,150305342,KowalskAP19,KowalskiQIP20,FonsecaRomeroPRA04,190709349,200309170,210308982,211113477}. 
In this paper we go beyond the paradigm of linear CPTP maps to design single-qubit gates that increase the length of the Bloch vector without changing its direction. It is well known that this operation can be implemented using linear Markovian completely positive trace-preserving (CPTP) channels, via the Gorini-Kossakowski-Sudarshan-Lindblad (GKSL) master equation \cite{GoriniJMP76,LindbladCMP76}. This provides a baseline which we call the {\bf linear CPTP} amplification gate.
The nonunital channel behind it is entropy decreasing, which is possible in an open system that compensates by producing enough environmental entropy as to not violate the second law. In addition to the linear CPTP gate, we also consider alternatives based on non-completely positive (non-CP) and nonlinear channels. The channels considered here are Markovian {\bf normalized PTP} channels taking the form $ X \mapsto \phi(X) / {\rm tr}[\phi(X)]$, where $\phi(X)$ is a continuous 1-parameter positive linear or nonlinear map satisfying ${\rm tr}[\phi(X)] \! \neq \! 0$ for all positive semidefinite (PSD) operators $X$. Normalized PTP channels fall into 4 classes, yielding 3 distinct forms of nonlinearity \cite{211105977}:

(i) {\it Linear PTP}: Linear $\phi$ and ${\rm tr}[\phi(X)]=1 \ {\rm for \  all} \ X$;

(ii) {\it NINO}: Linear $\phi$ and ${\rm tr}[\phi(X)] \neq 1 \ {\rm for \ some} \ X$; 

(iii) {\it State-dependent PTP}: Nonlinear $\phi$ and ${\rm tr}[\phi(X)] = 1 \ {\rm for \  all} \ X$;

(iv) {\it General normalized PTP}: Nonlinear $\phi$ and ${\rm tr}[\phi(X)] \neq 1 \ {\rm for \  some} \ X$. 

\noindent The linear CPTP gate belongs to class (i). A non-CP gate from class (i) will also be considered. Class (ii) leads to a restricted form of nonlinearity, where a diagonal nonlinear term is added to the master equation to conserve trace. This type of evolution equation extends a pure-state nonlinear Schr\"odinger equation first introduced by Gisin   \cite{GisinJPA81} in 1981, to mixed states 
\cite{BrodyPRL12,14057165,KowalskAP19,200309170,211105977}. Rembieli\'nski and Caban \cite{190603869} recently argued that this type of nonlinearity is causal (does not support superluminal signaling) and should not be excluded from a fundamental theory. We call these channels {\bf nonlinear in normalization only} (NINO) to emphasize their restricted form of nonlinearity. In Sec.~\ref{linear and nino section}, several amplification gates based on NINO channels are investigated. Class (iii) channels include unitary mean field theories such as the Gross-Pitaevskii equation for weakly interacting bosons, and they support Bloch-ball torsion, believed to be a powerful computational resource \cite{MielnikJMP80,PhysRevLett.81.3992,BechmannPLA98,0502072,150706334,211105977}. 
Our main result is a non-CP gate from class (i) and 
we do not discuss channels from class (iii) or (iv) in this paper.

 Quantum channels that are positive but not completely positive, or non-CP, are well known in open systems theory \cite{PechukasPRL94,ShajiPLA05,CarteretPRA08,13120908,150305342}, and are used to detect entanglement \cite{PeresPRL96}.
However the question of whether  non-CP channels could provide a {\it computational} advantage over linear CPTP channels appears to be largely unexplored \cite{150305342}, although they  have been shown to increase channel capacity \cite{0403092,0409026,0507045} in communication settings. Here we find an advantage for Bloch vector amplification, also called repolarization \cite{150305342}, and propose that an experimental demonstration of ``fast'' amplification would constitute a striking demonstration of a physical non-CP map.

\section{PSD cone}

\begin{figure}
\includegraphics[width=9.0cm]{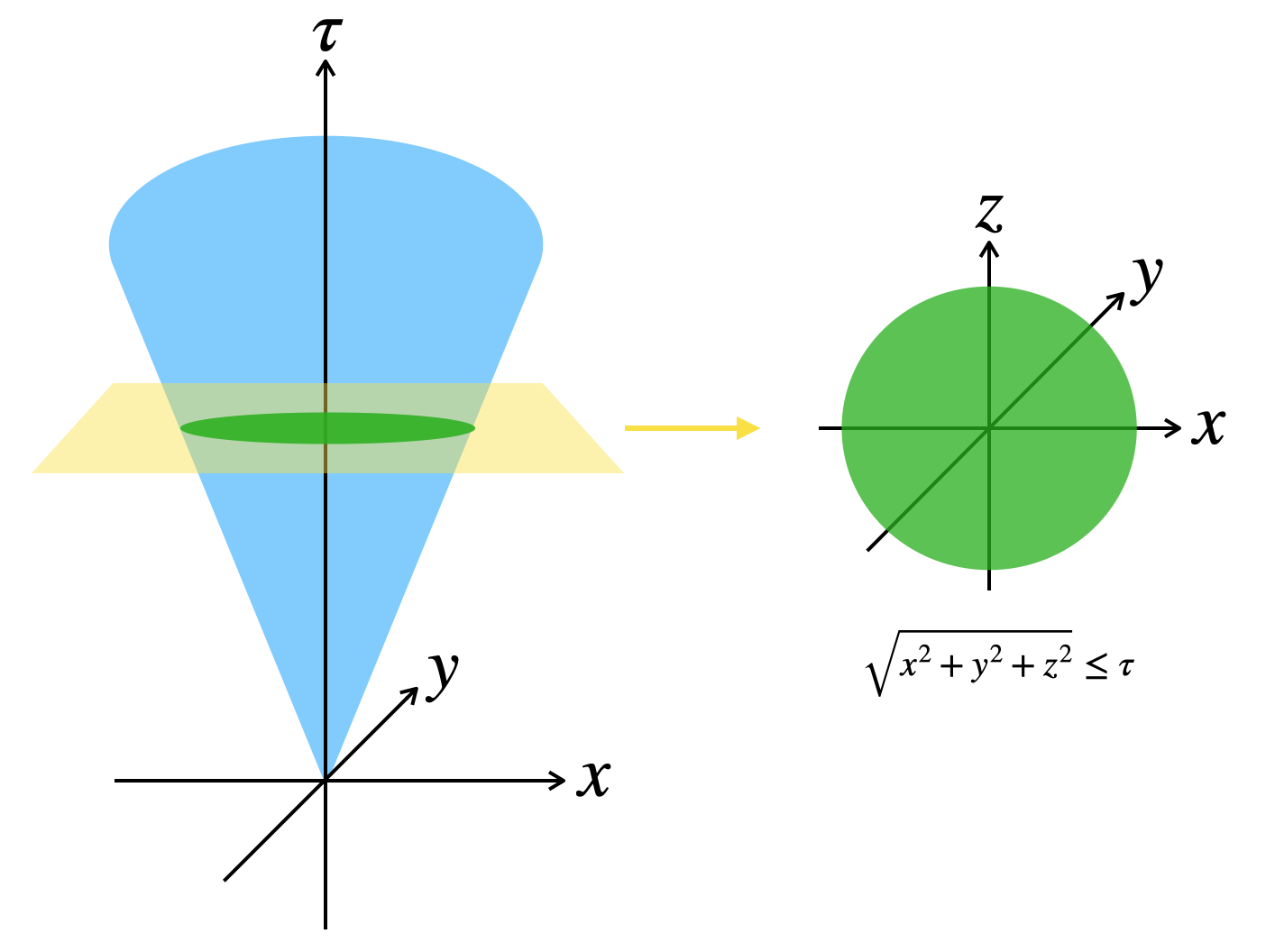} 
\caption{Extended state space of a qubit. On the left, the subspace with fixed trace is shown as a green circle, but it is really a Bloch ball with radius $\tau$.}
\label{psd cone figure}
\end{figure}

For the analysis of linear qubit channels it is sufficient to take, as the state space of a qubit, the Bloch sphere or ball, and to study the dynamics within that space. Here we will work in the larger space of PSD operators $X \succeq 0$ with strictly positive trace $\tau := {\rm tr}(X)$, a convex but noncompact set called the {\bf PSD cone}.\footnote{The PSD condition $X \succeq 0$ implies ${\rm tr}(X) \ge 0$. We further require that ${\rm tr}(X) \neq 0$, excluding the state $X=0$ at the apex of the cone.} Allowing density matrices to have a trace differing from the canonical value $\tau \! = \! 1$ is a straightforward extension of pure state quantum mechanics with square-integrable but unnormalized wave functions, and is equivalent to the canonical formulation as long as expectation values $\langle A \rangle := {\rm tr}(XA) / {\rm tr}(X)$ are properly defined, and ${\rm tr}(X) \neq 0$. There is little benefit to using the PSD cone representation with linear PTP channels due to their property of conserving $\tau$ for any initial value (this condition is part of their definition). However nonlinear PTP channels allow for a more restricted implementation of trace preservation, where the trace is conserved only if the initial trace has the canonical value $\tau \! = \! 1$. The PSD cone representation elucidates the mechanism of trace conservation in these cases. Let $X : {\cal H} \rightarrow {\cal H} $ be a linear operator on the system Hilbert space ${\cal H} = ( {\rm span}\{ | e_i \rangle \}_{i=1}^\DIM ,  \langle x|y\rangle )$, with complete orthonormal basis $ \{ | e_i \rangle \}_{i=1}^\DIM$ and inner product $ \langle x | y \rangle = \sum_{i=1}^\DIM x_i^* y_i$, and let $X^\dagger$ denote the adjoint of $X$ with respect to $\langle x | y \rangle$.
$x^*$ denotes complex conjugation and $I_\DIM$ is the $\DIM \! \times \! \DIM$ identity.  The set of bounded linear operators form a complex vector space $\BLO$. Let $\HER = \{ X \in \BLO : X = X^\dagger \} $ and $\PSD = \{ X \in {\rm Her}({\cal H})   : X \succeq 0 \}$ be the subsets of self-adjoint and PSD operators, respectively. In the qubit case, $N=2$, any $X \in \HER$ can be written in the Pauli basis as 
\begin{eqnarray}
X =  \frac{\tau I_2 + r^a \sigma^a}{2} = \frac{1}{2}
\begin{pmatrix}
\tau \! + \!  z & x \! - \!  iy \\
x \!  + \!  iy & \tau \! - \!  z & \\
\end{pmatrix} \! , \ \ 
{\rm spec}(X) = \frac{\tau \pm \sqrt{x^2 + y^2 + z^2} }{2}.
\label{extended pauli basis}
\end{eqnarray}
Here $\tau, \, r^a \in {\mathbb R}$, $a \in \{1,2,3\}$, and ${\rm spec}(X)$ contains the eigenvalues. The conditions for $X$ to be in $\PSD$ are (i) $\tau \ge 0$ and (ii) $| {\bf r}| = \sqrt{x^2 + y^2 + z^2} \le \tau$. For each $\tau >0$, the vector ${\bf r} = (x, y, z)$ must lie within a ball of radius $\tau$ centered at ${\bf r} = 0$,  defining a cone in ${\mathbb R}^4$ oriented along $\tau$. A 3d representation of this state space is given in Fig.~\ref{psd cone figure}.  

The physical interpretation of the Bloch vector is slightly different in the PSD cone picture. Namely, $ r^a = {\rm tr}( \sigma^a X)$ is equal to the expectation $\langle \sigma^a \rangle = {\rm tr}( \sigma^a X) / {\rm tr}(X) $ of Pauli matrix $\sigma^a $ times $\tau$:
\begin{eqnarray}
r^a  = {\rm tr}( \sigma^a X) =  \langle \sigma^a \rangle \tau.
\end{eqnarray}
The condition defining pure states is also modified. Let $\rho = X/\tau = \rho^2$ be a canonically normalized pure state. Therefore $X$ is pure if and only if
\begin{eqnarray}
X^2 = {\rm tr}(X) \, X  = \tau \, X.
\end{eqnarray}
Using  (\ref{extended pauli basis}) we see that these pure states lie on the surface of the PSD cone $| {\bf r} | = \tau$.
     
\clearpage

\section{Linear and NINO channels}
\label{linear and nino section}

In this paper we examine Bloch vector amplification gates based on the following NINO model
\cite{BrodyPRL12,14057165,KowalskAP19,200309170,211105977}:
\begin{eqnarray}
\frac{dX}{dt} = \{ L_{+}, X \}  +  \sum_\alpha \zeta_\alpha  B_\alpha X B_\alpha^\dagger +   g \, {\rm tr}(X\Omega) X, \ \ 
 \frac{d \tau}{dt} = ( g \tau \! - \! 1) \, {\rm tr}(X \Omega), \ \ 
 \tau = {\rm tr}(X), 
 \label{nino master equation}
 \end{eqnarray}
\begin{eqnarray}
\Omega := - 2 L_{+}   -    \sum_\alpha \zeta_\alpha \, B_\alpha^\dagger B_\alpha .
\label{omega definition}
\end{eqnarray}
Here $ X \succeq 0$ represents a (possibly unnormalized) qubit state. Time evolution consists of a smooth part plus random discontinuous jumps. The linear infinitesimal generator $L$ of the smooth part has been decomposed into Hermitian and anti-Hermitian components: 
$L_{\pm} = (L \pm L^\dagger)/2$. Equivalently, we can say that $ H = i L$ is the qubit Hamiltonian, which is non-Hermitian when $L_{+} \neq 0$. We expand the dissipative part $L_{+}$ in the Pauli basis as
\begin{eqnarray}
L_{+} = \ell_\mu \sigma^\mu, \ \  \sigma^\mu = (I_2, \,  \sigma^1, \sigma^2, \sigma^3), \ \ \ell_\mu \in {\mathbb R}^4.
 \label{L+ expansion}
 \end{eqnarray}
The anti-Hermitian part $L_{-}$ generates unitary evolution. Because amplification is purely nonunitary, we set $L_{-}$ to zero. The $B_\alpha \in \BLO$ are a set of linearly independent jump operators. The  $\zeta_\alpha = \pm 1$ are signs of the Choi matrix eigenvalues  (all nonnegative for CPTP channels). Any jump operator with $\zeta_\alpha = -1$ indicates a non-CP channel \cite{PechukasPRL94,ShajiPLA05,CarteretPRA08,13120908}. $g \in {\mathbb R}$ controls the strength of the nonlinear term. The observable $\Omega \in \HER$ is chosen to conserve trace and plays an important role in NINO channels because it governs the dynamics in the $\tau$ direction of the PSD cone.

The trace equation in (\ref{nino master equation}) shows that there are two distinct ways to achieve trace-conservation $d\tau/dt = 0$:  The first is the linear option, $g=0$, which requires $\Omega=0$ and leads to the GKSL equation and the linear CPTP gate (if the $\zeta_\alpha$ are positive). The linear option conserves trace for any initial $\tau$. The second option is to make use of the nonlinearity and fix $g = 1$, leading to the NINO gates. In this case
\begin{eqnarray}
 \frac{d \tau}{dt} =  (\tau - 1) \, {\rm tr}(X \Omega) .
 \label{g=1 nino trace equation}
 \end{eqnarray}
 This conserves trace if  $\tau$ starts with the canonical value $\tau = 1$. The $\tau = 1$ plane in the PSD cone is a fixed plane of the channel. The fixed plane is locally stable wherever ${\rm tr}(X \Omega) \le 0$. The operator $\Omega$ can lead to an intricate fixed point structure in the PSD cone, including instabilities in dynamically {\it inaccessible} regions of the cone that nevertheless leave their imprint on the accessible regions in an intuitive way. 

The NINO model (\ref{nino master equation}) has a continuous symmetry that is absent (pushed to infinity) in the linear model: Under
\begin{eqnarray}
L_{+} \mapsto L_{+}  + c  I_2,  \ c \in R
 \label{shift}
 \end{eqnarray}
we have 
\begin{eqnarray}
\Omega \mapsto \Omega  -2  c  I_2
 \label{omega shift}
 \end{eqnarray}
 and
\begin{eqnarray}
 \frac{dX}{dt} \mapsto \frac{dX}{dt} + 2 c (1-g\tau)X,
 \label{nino symmetry}
 \end{eqnarray}
so the equation of motion is invariant if 
\begin{eqnarray}
g  \tau = 1.
\label{nino g condition}
\end{eqnarray}
The condition (\ref{nino g condition}) is the same as that for trace conservation in (\ref{nino master equation}). Note that the jump operators do not change under this transformation. In certain cases this symmetry can be used to map a NINO channel to a dual linear channel (Sec.~\ref{pseudoliner nino section}).
  
We will consider a sequence of increasingly complex NINO channels and amplification gates constructed from a set of jump operators $\{ B_0,  B_1, B_2, B_3 \}$ listed in Table \ref{jump operator table}, combined with specific values of $L_{+}$. By amplification we mean a process that smoothly increases $|{\bf r}|$ from 0 to $\tau$. Without loss of generality we can amplify along the $x$ axis of the Bloch ball. The first jump operator $B_0$ in Table \ref{jump operator table} is chosen to produce $x$-axis amplification in the linear CPTP limit (when combined with an appropriate $L_{+}$). The additional jump operators allow for increased control over the fixed points of the map. The amplification gates assume that the qubit is initially prepared in the state $X=\frac{I_2}{2}$, so they require a nonunital channel or an unstable fixed point. Note that 
\begin{eqnarray}
\bigg( \! \frac{dX}{dt} \! \bigg)_{\! \! \! \frac{I_2}{2} } \! 
=  L_{+} +   \frac{1}{2} \displaystyle  \sum_\alpha \zeta_\alpha B_\alpha B_\alpha^\dagger  
+ g \frac{ {\rm tr}(\Omega) }{4} \,  I_2
=  \frac{1}{2}   \sum_\alpha \zeta_\alpha  [B_\alpha, B_\alpha^\dagger]   - \frac{\Omega}{2} 
+ g \frac{ {\rm tr}(\Omega) }{4} \,  I_2.
\label{initial velocity}
\end{eqnarray}
In the linear theory, $g \! = \! \Omega \! = \! 0$, this expression recovers the well known result that a nonunital Markovian channel requires one or more nonnormal jump operators. However in a NINO channel with $\Omega \neq 0$ we can implement nonunital maps with normal jump operators or even with no jump operators (Sec.~\ref{no-jump nino section}).

\begin{table}[htb]
\centering
\caption{Jump operators used in this paper. The $\sigma^1, \sigma^2, \sigma^3 $ are Pauli matrices. The $m \in {\mathbb R}$ are constants determining mean jump frequencies, which we assume to be individually controllable.}
\begin{tabular}{|c|c|c|c|c|c|c|}
 \hline
$\alpha$ & $B_\alpha$ & $B_\alpha^\dagger B_\alpha$ & $B_\alpha B_\alpha^\dagger$ & $\xi_\alpha$ & $G_\alpha $ & $C_\alpha$ \\
 \hline
0 &  $m (\sigma^2 + i \sigma^3)$ &  $2 m^2 ( I_2 - \sigma^1)$ & $2 m^2 ( I_2 + \sigma^1)$  & $m(0,0,1,i)$ & $ m^2 \!  \begin{pmatrix} -2 & 0 & 0 \\ 0 & 0 & 0 \\ 0 & 0 & 0 \\ \end{pmatrix}$ &  $ m^2  \! \begin{pmatrix} 2  \\ 0 \\  0 \end{pmatrix}$ \\
 \hline
1 &   $m (\sigma^1 + \sigma^2)$  &  $2 m^2  I_2$  & $2 m^2  I_2$ & $m(0,1,1,0)$ & $ m^2 \!  \begin{pmatrix} 0 & 2 & 0 \\ 2 & 0 & 0 \\ 0 & 0 & -2 \\ \end{pmatrix}$ &  0  \\
 \hline
 2 &   $ m (I_2 + \sigma^3) $  &  $2 m^2 ( I_2 + \sigma^3)$ & $2 m^2 ( I_2 + \sigma^3)$ & $m(1,0,0,1)$  & $ m^2  \!  \begin{pmatrix} 0 & 0 & 0 \\ 0 & 0 & 0 \\ 0 & 0 & 2 \\ \end{pmatrix} $ &  $ m^2  \! \begin{pmatrix} 0  \\ 0 \\  2 \end{pmatrix}$ \\
 \hline
 3 &  $m \sigma^3$ & $ m^2 I_2 $ & $ m^2 I_2 $ &   $m(0,0,0,1)$ & $  m^2  \!  \begin{pmatrix} -1 & 0 & 0 \\ 0 & -1 & 0 \\ 0 & 0 & 1 \end{pmatrix} $ &  0 \\
 \hline
 \end{tabular}
\label{jump operator table}
\end{table}

In the Pauli basis (\ref{nino master equation}) becomes
\begin{eqnarray}
\frac{d r^a}{dt} &=& {\rm tr} \bigg( \! \sigma^a \frac{dX}{dt}\bigg) =
G_{L_+}^{ab} r^b + C_{L_+}^a \tau + 
\sum_\alpha  \zeta_\alpha \bigg( \! G_\alpha^{ab} r^b+ C_\alpha^a \tau \! \bigg) + g \, {\rm tr}(X \Omega) \, r^a \! .
\label{pauli basis nino master equation}
\end{eqnarray}
The first and second terms on the right side of (\ref{pauli basis nino master equation}) come from the $ \{ L_{+} , X \} $ term in (\ref{nino master equation}). Here $G_{L_+}= {\rm tr} (L_{+}) I_3 $ and $C_{L_+}^a = {\rm tr}(\sigma^a L_{+}) = 2 \ell_a$. The summation over $\alpha$ in (\ref{pauli basis nino master equation}) contains the contributions 
\begin{eqnarray}
G_\alpha^{ab}  =  {\rm tr} ( \sigma^a  B_\alpha \sigma^b B_\alpha^\dagger )/2  \ \ {\rm and} \ \ 
C_\alpha^a =  {\rm tr} ( \sigma^a  B_\alpha B_\alpha^\dagger )/2 
\end{eqnarray}
from one or more jump operators. Expanding any $B \in {\mathbb C}^{2 \times 2}$ in complex coordinates
\begin{eqnarray}
B = \xi_\mu \sigma^\mu, \ \  \sigma^\mu = (I_2, \,  \sigma^1, \sigma^2, \sigma^3), \ \ \xi_\mu \in {\mathbb C}^4
\end{eqnarray}
leads to
$ G^{ab}_\alpha =  \big( |\xi_0|^2 - |\xi_1|^2 -  |\xi_2|^2  -  |\xi_3|^2 \big)  \delta^{ab} 
+ 2  \,  {\rm Im} (\xi^*_0 \xi_c ) \varepsilon^{abc} 
+ 2  \,  {\rm Re} (\xi^*_a \xi_b ) $.
The resulting $G_\alpha$ matrices for each jump operator are given in Table \ref{jump operator table}. The last term in (\ref{pauli basis nino master equation})
containing ${\rm tr}(X \Omega)  =  (\tau \,  {\rm tr}(\Omega) + {\bf r} \cdot {\rm tr}({\bm \sigma}  \Omega) ) /2  $
 is diagonal and nonlinear. The vectors $C_{L_+} , C_\alpha \in {\mathbb R}^3$ determine the initial velocity $dX/dt$, nonunitality, and fixed points of the channel. These are also given in Table \ref{jump operator table}.

\subsection{Linear CPTP}

If $g=0$, the trace equation in (\ref{nino master equation}) requires  that $\Omega = 0$, leading to the linear channel
\begin{eqnarray}
 \frac{dX}{dt}  =   \{ L_{+}, X \} +  \sum_\alpha \zeta_\alpha  B_\alpha X B_\alpha^\dagger , \ \
\frac{d \tau}{dt}  \!  =  0, \  \
L_{+} = - \frac{1}{2}  \sum_\alpha \zeta_\alpha B_\alpha^\dagger B_\alpha  .
\label{general gksl master equation}
\end{eqnarray}
If the $\zeta_\alpha = 1$ then (\ref{general gksl master equation})  reduces to the GKSL master equation and the map is CPTP. Here we will consider a model with a single jump operator $B_0$ from 
Table \ref{jump operator table} with $\zeta_0 = 1$:
\begin{eqnarray}
 \frac{dX}{dt}  =   \{ L_{+}, X \} +  B_0 X B_0^\dagger , \ \
\frac{d \tau}{dt}  \!  =  0, \  \
L_{+} = - \frac{B_0^\dagger B_0 }{2}= m^2 (\sigma^1 - I_2) .
\label{gksl master equation}
\end{eqnarray}
The requirement $\Omega = 0$ imposes conditions $\ell_0 = -m^2$, $\ell_1 = m^2$, $\ell_2 = 0$, and $\ell_3 = 0$ on $L_{+}$. In the Pauli basis
\begin{eqnarray}
\frac{d r^a}{dt} = G^{ab} r^b + C^a  \tau, \ \ 
G = G_{L_+}+ G_0 = m^2  \! 
\begin{pmatrix}
 -4  & 0 & 0 \\
0  &  -2    & 0  \\
 0 & 0 &  -2 
\end{pmatrix} \! , \ \ 
C = C_{L_+} + C_0  = m^2  \! 
\begin{pmatrix}
4 \\
0  \\
0 
\end{pmatrix} \! , \ \ \ \ \ \ 
\label{gksl master equation pauli basis}
\end{eqnarray}
leading to the equations of motion
\begin{eqnarray}
\frac{dx}{dt} &=& 4 m^2 (\tau-x), \\
\frac{dy}{dt} &=& -2 m^2 y, \\
\frac{dz}{dt} &=& -2 m^2 z .
\label{gksl equations}
\end{eqnarray}
The solution giving the desired Bloch vector amplification gate is
\begin{eqnarray}
 x(t) = (1 - e^{-4 m^2 t}) \tau  , \ \ 
 x(0) = 0, \ \ 
x(\infty) =  \tau, \ \ y, z=0.
\label{gksl solution}
\end{eqnarray}
The channel has a single stable fixed point at ${\bf r}^{\rm fp} = (\tau, 0, 0 )$, but the velocity decreases as this fixed point is approached. Due to this critical slowing down, it takes infinitely long to reach the pure state at $x=\tau$, although it becomes exponentially close for $t \gg 1/4 m^2$. To quantify the critical slowing down, let $\delta = \tau - x \ge 0$. Thus $dx/dt$ vanishes linearly with $ \delta$ at the fixed point $\delta =0$. 

\subsection{No-jump NINO}
\label{no-jump nino section}

The first NINO gates we consider are variations on the linear CPTP gate. The simplest is an amplification gate is based on (\ref{nino master equation}) with $g=1$ and no jump operators. The single model parameter is $L_{+}$, which we take to be  $L_{+} = \ell_0 I_2 + \ell_1 \sigma^1$, $ \ell_0,  \ell_1 \in {\mathbb R}$, as in (\ref{gksl master equation}). The equation of motion for 
the {\bf no-jump NINO} model is 
\begin{eqnarray}
 \frac{dX}{dt} = \{ L_{+}, X \}  +  {\rm tr}(X \Omega ) X, \ \ 
 \frac{d \tau}{dt} =  (\tau - 1) \, {\rm tr}(X \Omega), \ \ 
\Omega = \! - 2 L_{+} \! = \! -2(\ell_0 I_2 \!  + \!  \ell_1 \sigma^1) , \ \ 
\label{nojump nino master equation}
\end{eqnarray}
a nonlinear channel with a purely non-Hermitian Hamiltonian $H = i L_{+}$ and no jump operators. In the Pauli basis,
\begin{eqnarray}
\frac{d r^a}{dt} = G^{ab} r^b + C^a  \tau, \ \ 
G = 2 \ell_0 I_3 + {\rm tr} (X\Omega) I_3,
\ \ C = \ell_1 \! 
\begin{pmatrix}
2 \\
0 \\
0
\end{pmatrix} \! . \ \ 
\end{eqnarray}
The equations of motion are
\begin{eqnarray}
\frac{dx}{dt} &=& 2 \ell_0 (1-\tau)x  + 2 \ell_1 (\tau - x^2), \\
\frac{dy}{dt} &=& 2 \ell_0 (1-\tau)y - 2 \ell_1  \, xy , \\
\frac{dz}{dt} &=& 2 \ell_0 (1-\tau)z - 2 \ell_1  \, xz .
\end{eqnarray}
In the $\tau = 1$ plane,
\begin{eqnarray}
\frac{dx}{dt} = 2 \ell_1 (1 - x^2), \ \ 
\frac{dy}{dt} = - 2 \ell_1  \, xy , \ \ 
\frac{dz}{dt} = - 2 \ell_1  \, xz .
\end{eqnarray}
Now there are two fixed points at ${\bf r}^{\rm fp}_{\pm} = (\pm 1, 0, 0 )$. ${\bf r}^{\rm fp}_{+}$ is stable but ${\bf r}^{\rm fp}_{-}$ is not. Relative to the linear CPTP gate, the no-jump NINO gate adds a second fixed point at ${\bf r}^{\rm fp}_{-}$,  but does not seem to offer any computational benefit. The velocity again decreases linearly in $\delta = \tau - x \ge 0$ as the fixed point is approached.

\subsection{One-jump NINO}

Next we consider a NINO gate using the jump operator $B=B_0$ from the linear CPTP gate, but with $L_{+} = 0$ and trace conserved nonlinearly. The equation of motion for the {\bf one-jump NINO} channel is
\begin{eqnarray}
 \frac{dX}{dt} =  B X B^\dagger + {\rm tr}(X \Omega) X, \ \ 
\frac{d \tau}{dt} =  (\tau - 1) \, {\rm tr}(X \Omega), \ \ 
\Omega =  - B^\dagger B 
= 2 m^2 (\sigma^1 - I_2).
\label{onejump nino master equation}
\end{eqnarray}
\begin{eqnarray}
\frac{d r^a}{dt} = G^{ab} r^b + C^a  \tau, \ \ 
G = 
m^2 
\begin{pmatrix}
-2 & 0 & 0 \\
0  & 0 & 0 \\
0  & 0 & 0 
\end{pmatrix}
+  {\rm tr} (X\Omega) I_3,
\ \ C = 
m^2 
\begin{pmatrix}
2   \\
0 \\
0
\end{pmatrix} \! . \ \ 
\end{eqnarray}
In the $\tau = 1$ plane,
\begin{eqnarray}
\frac{dx}{dt} = 2 m^2  (x-1)^2, \ \ 
\frac{dy}{dt} =  2 m^2  \, (x-1) y , \ \ 
\frac{dz}{dt} =  2 m^2 \, (x-1)z ,
\label{onejump nino odes}
\end{eqnarray}
which has a single fixed point at ${\bf r}^{\rm fp} = (1, 0,0)$. To determine its stability, let $\delta = \tau - x \ge 0$ and rewrite (\ref{onejump nino odes}) as
\begin{eqnarray}
\frac{dx}{dt} = 2 m^2  \delta^2, \ \ 
\frac{dy}{dt} = - 2 m^2 \delta  \, y   , \ \ 
\frac{dz}{dt} = - 2 m^2 \delta   \, z  .
\label{onejump nino odes expanded}
\end{eqnarray}
Inside the PSD cone, $\delta > 0$ and the motion is stable, but the critical slowing down is worse now because $dx/dt$ vanishes as $\delta^2$ at the fixed point. Outside the PSD cone, however, the motion is unstable, and $x$ increases without bound. The unstable region does not appear to be accessible dynamically when starting from the initial state $X  = \frac{I_2}{2}$. 

\clearpage

\subsection{Pseudo-liner NINO}
\label{pseudoliner nino section}

Next we consider a special subclass of NINO channels with $g = 1$ and $\Omega = \kappa I_2$ proportional to the identity. In this case the observable $\Omega$ vanishes except for a component in the $I_2$ direction. The equation of motion for 
the {\bf pseudo-liner NINO} channel is 
\begin{eqnarray}
\frac{dX}{dt} = \{ L_{+}, X \}  +  \sum_\alpha \zeta_\alpha  B_\alpha X B_\alpha^\dagger +   g \kappa \tau X, \ \ 
 \frac{d \tau}{dt} = (  g \tau \! - \! 1) \kappa \tau , \ \ g=1.
 \label{pseudo linear master equation}
 \end{eqnarray}
Here the operators $L_{+} \in \HER$ and $B_\alpha \in \BLO$ are no longer independent; they are constrained to satisfy
\begin{eqnarray}
\Omega = - 2 L_{+}   -    \sum_\alpha \zeta_\alpha \, B_\alpha^\dagger B_\alpha = \kappa I_2, \ \ \kappa \in {\mathbb R}.
 \label{pseudo linear omega condition}
\end{eqnarray}
The condition $g=1$ is required by trace conservation (applied to a $\tau = 1$ initial state).
In this subclass the nonlinearity is effectively invisible, because 
\begin{eqnarray}
g \, {\rm tr}(X \Omega)  X = g \kappa  \tau X ,
\label{invisibility condition}
\end{eqnarray}
which acts linearly when restricted to the $\tau =1$ plane. This implies a duality between pseudo-linear NINO channels and linear PTP channels. Assuming  (\ref{pseudo linear master equation}) and
(\ref{pseudo linear omega condition}), the dual model is obtained by
\begin{eqnarray}
L_{+} \mapsto L_{+}  + \frac{\kappa}{2} I_2, \ \ 
g \mapsto 0.
 \label{duality map}
\end{eqnarray}
We should think of (\ref{duality map}) as being composed of two physically distinct steps. In the first step we shift 
$L_{+} \mapsto L_{+}  + \frac{\kappa}{2} I_2$ while keeping $g=1$, which [according to (\ref{omega shift})] rescales $\Omega \mapsto 0$. It is important that $g=1$ during this first step, as required by (\ref{nino g condition}). However now that $\Omega = 0$, the trace can be conserved linearly, while violating condition (\ref{nino g condition}). So in the second step we switch off the nonlinearity.
 
\clearpage
 
Consider the following example of a pseudo-linear NINO channel for a qubit: Combine the jump operator $B_0$ from the linear CPTP gate with $L_{+} = \ell_1 \sigma^1$, where $\ell_1 = m^2$. Then $\Omega = \kappa I_2$ with $\kappa = -2 m^2$. In the Pauli basis
\begin{eqnarray}
& {\displaystyle \frac{d r^a}{dt} } = G^{ab} r^b + C^a  \tau, &  \\
&G = G_0 
- 2 m^2 \tau I_3 = 
m^2 
\begin{pmatrix}
-2 -2 \tau & 0 & 0 \\
0  & -2 \tau & 0 \\
0  & 0 & -2 \tau 
\end{pmatrix}, & \\
&C = C_{L_+} + C_0 = m^2 
\begin{pmatrix}
4   \\
0 \\
0
\end{pmatrix} .&
\end{eqnarray}
In the $\tau = 1$ plane this leads to
\begin{eqnarray}
\frac{dx}{dt} = 4 m^2 (1-x), \ \ 
\frac{dy}{dt} = -2 m^2 y, \ \ 
\frac{dz}{dt} = -2  m^2 z,
\end{eqnarray}
with same solution (\ref{gksl solution}) as the linear CPTP gate. The nonlinearity in the pseudo-liner model is invisible in the Bloch ball picture but is apparent in the PSD cone, where it produces a fixed plane at  $\tau = 1$. The fixed plane is lineary stable if $m \neq 0$.  This form of nonlinearity does not appear to offer any computational advantage over the linear CPTP gate, as expected from the duality (\ref{duality map}).

\subsection{Three-jump NINO}
\label{three-jump nino section}

Here we examine a three-jump NINO channel with a particularly striking fixed point structure due to the presence of fixed {\it lines} in the PSD cone. These fixed lines, when properly controlled, enable fast Bloch vector amplification without deceleration. 
The equation of motion for the {\bf three-jump NINO} channel is
\begin{eqnarray}
\frac{dX}{dt} = \{ L_{+}, X \}  +  \sum_\alpha \zeta_\alpha  B_\alpha X B_\alpha^\dagger +   g \, {\rm tr}(X\Omega) X, \ \ 
 \Omega = - 2 L_{+}   -    \sum_\alpha \zeta_\alpha \, B_\alpha^\dagger B_\alpha .
 \label{three jump master equation}
 \end{eqnarray}
 Here the sum is over $\alpha =1,2,3$, with Choi eigenvalue signs $\zeta_1 = 1$,  $\zeta_2 = 1$,  $\zeta_3 = -1$. The jump operators $B_1$,  $B_2$, $B_3$ are given in Table \ref{jump operator table}. In addition to these three jump operators we include a dissipative part $L_{+} = \ell_\mu \sigma^\mu, \ \sigma^\mu = (I_2, \,  \sigma^1, \sigma^2, \sigma^3), \ \ell_\mu \in {\mathbb R}^4$. 
 We do not include jump operator $B_0$ in this gate. $g \in {\mathbb R}$ is arbitrary. In the Pauli basis,
\begin{eqnarray}
 \frac{d r^a}{dt} = G^{ab} r^b + C^a  \tau, \ \ G = G_{L_+}+ {\displaystyle  \sum_{\alpha=1}^3  } \zeta_\alpha G_\alpha +  g \, {\rm tr}(X\Omega) I_3, \ \ C = C_{L_+} + {\displaystyle  \sum_{\alpha=1}^3  } \zeta_\alpha C_\alpha .
\end{eqnarray}
Then
\begin{eqnarray}
G &=& {\rm tr}(L_{+}) I_3 + 
m_1^2 
\begin{pmatrix}
0  & 2 & 0 \\
2 & 0 & 0 \\
0  & 0 & -2 
\end{pmatrix}
+ m_2^2
\begin{pmatrix}
0  & 0 & 0 \\
0 & 0 & 0 \\
0  & 0 & 2 
\end{pmatrix}
- m_3^2
\begin{pmatrix}
-1  & 0 & 0 \\
0 & -1 & 0 \\
0  & 0 & 1 
\end{pmatrix}
+ g \, {\rm tr}(X\Omega) I_3,  \\
C &=& C_{L_+} + C_2 =
\begin{pmatrix}
2 \ell_1   \\
2 \ell_2 \\
2 \ell_3
\end{pmatrix}
+ m_2^2 
\begin{pmatrix}
0   \\
0 \\
2
\end{pmatrix} \! .
\end{eqnarray}
Next we set the jump operator strengths to be
\begin{eqnarray}
 m_1 = \sqrt{M/2} , \ \ m_2 =  \sqrt{M/2}, \ \   m_3 =  \sqrt{M - {\textstyle \frac{\Gamma}{2}}} ,
\label{jump operator settings}
\end{eqnarray}
where 
\begin{eqnarray}
 M \ge \frac{\Gamma}{2} \ge 0.
\end{eqnarray}
With these settings 
\begin{eqnarray}
&& G = {\rm tr}(L_{+}) I_3 + 
\begin{pmatrix}
0  & M & 0 \\
M & 0 & 0 \\
0  & 0 & 0 
\end{pmatrix}
+
\begin{pmatrix}
M - {\textstyle \frac{\Gamma}{2}}  & 0 & 0 \\
0 & M - {\textstyle \frac{\Gamma}{2}} & 0 \\
0  & 0 & {\textstyle \frac{\Gamma}{2}} - M
\end{pmatrix}
+ g \, {\rm tr}(X\Omega) I_3,  \ \ 
\\
&& C = 
\begin{pmatrix}
2 \ell_1   \\
2 \ell_2 \\
2 \ell_3
\end{pmatrix}
+ 
\begin{pmatrix}
0   \\
0 \\
M
\end{pmatrix} \! , \\
&& \sum_{\alpha=1}^3 \zeta_\alpha  B_\alpha^\dagger B_\alpha  
= 2 m_1^2 I_2 + 2 m_2^2(I_2 + \sigma^3) - m_3^2 I_2
= (M + {\textstyle \frac{\Gamma}{2}} ) I_2 + M \sigma^3 , \\
&& \Omega = -2 L_{+} -  (M + {\textstyle \frac{\Gamma}{2}} ) I_2 
-  M  \sigma^3 .
\end{eqnarray}

Next we choose $L_{+} = -(M/2) \sigma^3$ to tune  the $\sigma^3 $ component of $\Omega$ to zero, leading to a pseudo-linear channel:
\begin{eqnarray}
\Omega = - (M + {\textstyle \frac{\Gamma}{2}}) I_2 , \ \ 
{\rm tr}(X\Omega) = - (M + {\textstyle \frac{\Gamma}{2}}) \tau .
\label{three-jump nino pseudolinear setting}
\end{eqnarray}
According to (\ref{initial velocity}), however, when $g=1$ the resulting channel satisfies 
\begin{eqnarray}
\bigg( \! \frac{dX}{dt} \! \bigg)_{\! \! \! \frac{I_2}{2} } \! = 0
\end{eqnarray}
and is therefore unital. This would seem to preclude its use for Bloch vector amplification, but this conclusion does not apply if the fixed point is unstable. In the pseudo-linear case (\ref{three-jump nino pseudolinear setting}) we have
\begin{eqnarray}
G =  
\begin{pmatrix}
M - {\textstyle \frac{\Gamma}{2}}  & M & 0 \\
M & M - {\textstyle \frac{\Gamma}{2}}  & 0 \\
0  & 0 & {\textstyle \frac{\Gamma}{2}} - M
\end{pmatrix}
- \tau (M + {\textstyle \frac{\Gamma}{2}}) I_3, \ \ 
C = 0.
\end{eqnarray}
Finally, upon restriction to the $\tau = 1$ plane and  $g=1$ we obtain the following qubit equation of motion in the three-jump NINO model:
\begin{eqnarray}
 \frac{d r^a}{dt} = G^{ab} r^b, \ \ G = \begin{pmatrix}
-\Gamma & M & 0 \\
M & -\Gamma  & 0 \\
0  & 0 & -2M
\end{pmatrix} \! , \ \ 
M, \Gamma \in {\mathbb R}, \ \ 
 M \ge \frac{\Gamma}{2} \ge 0.
 \label{three-jump model pauli basis}
\end{eqnarray}

Let's examine the equations of motion for the model
(\ref{three-jump model pauli basis}), which is similar to a model investigated in \cite{211105977} but now without torsion:
\begin{eqnarray}
\frac{dx}{dt} = - \Gamma x + M y, \ \ 
\frac{dy}{dt} = - \Gamma y + M x ,\ \ 
\frac{dz}{dt} =  -2M z .
\label{three jump master odes}
\end{eqnarray}
If $M \neq \Gamma$ the channel has a single fixed point ${\bf r}^{\rm fp}_0 = (0,0,0)$ at the center of the Bloch ball. To examine its stability, switch to rotated coordinates
\begin{eqnarray}
\xi_{\pm} := \frac{y \pm x}{2}.
\end{eqnarray}
Then
\begin{eqnarray}
\frac{d \xi_{+} }{dt} = (M - \Gamma) \, \xi_{+} , \ \ 
\frac{d \xi_{-} }{dt} = - (M + \Gamma) \, \xi_{-} .
\end{eqnarray}
Note that if $M = \Gamma$, the entire $\xi_{+}$ axis (the line $y=x$, $z=0$) is a fixed {\it line} of the map. If $M \neq \Gamma$, there are no fixed lines and the $\xi_{+}$ axis flows toward or away from ${\bf r}^{\rm fp}_0$. The $\xi_{+}$ direction is stable for $M < \Gamma$ but is unstable for $M > \Gamma$. Therefore when $M > \Gamma$ the fixed point ${\bf r}^{\rm fp}_0$ becomes unstable. The instability of the fixed point at $X = \frac{I_2}{2}$ allows this unital channel to be used for Bloch vector amplification. The solution giving the desired amplification gate for $ M > \Gamma$ is
\begin{eqnarray}
x(t) = y(t) = e^{(M-\Gamma)t} - 1, \ \  x(t_{\rm gate} ) = y(t_{\rm gate} ) = \frac{\tau}{\sqrt 2} , \ \  
t_{\rm gate} = \frac{\log( 1 + \frac{\tau}{\sqrt 2})}{M-\Gamma} , \ \
z=0. \ \ 
\label{three-jump nino gate}
\end{eqnarray}
Note that the $\xi_{-}$ direction is always stable in this model. This is a great improvement over the linear CPTP gate because it does not decelerate.

An amplification gate based on the three-jump NINO model (\ref{three-jump model pauli basis}) requires a small modification due to the instability at the starting point $X = \frac{I_2}{2}$. The simplest modification would be to initialize the qubit slightly away from ${\bf r} = (0,0,0)$ before switching on the nonlinearity. This can be achieved by applying the linear CPTP amplification gate for a short duration to pre-amplify the state to ${\bf r} = (x,0,0)$ with small positive $x$. In the rotated frame the pre-amplified state is
\begin{eqnarray}
\xi_{+} = \frac{x}{2} , \ \ 
\xi_{-} = - \frac{x}{2} , \ \ 0 < x \ll 1.
\end{eqnarray}
Applying the three-jump NINO channel then successfully amplifies the qubit state. 

\subsection{Linear non-CP}
\label{linear non-cp section}

The final model we consider is motivated by the  three-jump gate (\ref{three-jump nino gate}), which supports fast Bloch vector amplification without deceleration. As explained above, the NINO model (\ref{nino master equation}) is invariant under a shift (\ref{shift}) of the dissipative part $L_{+}$ of the non-jump component of the linear infinitesimal generator.\footnote{Recall that $ i L_{+} $ is an anti-Hermitian but otherwise arbitrary qubit Hamiltonian. The Hermitian part of the Hamiltonian vanishes here because  amplification is nonunitary.} 
Furthermore, in the special case of a pseudo-linear NINO channel (Sec.~\ref{pseudoliner nino section}), where the observable $\Omega$ is proportional to the identity, the invariance leads to  the duality (\ref{duality map}) between pseudo-linear NINO channels with $g=1$ and strictly linear channels with $g=0$. Here we use this duality to construct a linear non-CP channel and gate equivalent to those of Sec.~\ref{three-jump nino section}, specifically to the pseudo-linear model (\ref{three-jump nino pseudolinear setting}). The equivalent {\bf linear non-CP}  model, an instance of (\ref{general gksl master equation}), is
\begin{eqnarray}
 \frac{dX}{dt}  =   \{ L_{+}, X \} +  \sum_\alpha \zeta_\alpha  B_\alpha X B_\alpha^\dagger , \ \
\frac{d \tau}{dt}  \!  =  0, \  \
L_{+} = - \frac{1}{2}  \sum_\alpha \zeta_\alpha B_\alpha^\dagger B_\alpha , \ \ g=0. 
\label{dual linear ptp master equation}
\end{eqnarray}
Here the sum is over $\alpha =1,2,3$, with signs $\zeta_1 = 1$, 
 $\zeta_2 = 1$,  $\zeta_3 = -1$, indicating 
a non-CP channel \cite{PechukasPRL94,ShajiPLA05,CarteretPRA08,13120908}. The jump operators $B_1$, 
$B_2$, $B_3$ are given in Table \ref{jump operator table} and are the same as in  Sec.~\ref{three-jump nino section}. In addition to these jump operators we include the $L_{+}$ specified in (\ref{dual linear ptp master equation}). Then 
\begin{eqnarray}
 \frac{d r^a}{dt} = G^{ab} r^b + C^a  \tau,\end{eqnarray}
where, after using (\ref{jump operator settings}),
\begin{eqnarray}
G = {\rm tr}(L_{+}) I_3 + 
\begin{pmatrix}
M-\frac{\Gamma}{2}   & M & 0 \\
M & M-\frac{\Gamma}{2}  & 0 \\
0  & 0 & \frac{\Gamma}{2} - M
\end{pmatrix} \! ,  \ \ 
C = 
\begin{pmatrix}
2 \ell_1   \\
2 \ell_2 \\
2 \ell_3
\end{pmatrix}
+ 
\begin{pmatrix}
0   \\
0 \\
M
\end{pmatrix} \! ,
\end{eqnarray}
where
\begin{eqnarray}
L_{+} = - \frac{ (M + \frac{\Gamma}{2}) I_2 + M \sigma^3 }{2}, \ \ {\rm tr}(L_{+}) =  -(M +   
{\textstyle \frac{\Gamma}{2 } } ) .
\end{eqnarray}
Then $ \ell_1 = \ell_2 = 0$ and $\ell_3  = - \frac{M}{2}$, leading to
\begin{eqnarray}
 \frac{d r^a}{dt} = G^{ab} r^b, \ \ G = \begin{pmatrix}
-\Gamma & M & 0 \\
M & -\Gamma  & 0 \\
0  & 0 & -2M
\end{pmatrix} \! , \ \ 
C = 0, \ \ 
M, \Gamma \in {\mathbb R}, \ \ 
 M \ge \frac{\Gamma}{2} \ge 0,
 \label{linear ptp model pauli basis}
\end{eqnarray}
as in (\ref{three-jump model pauli basis}).
This supports the fast amplification gate
(\ref{three-jump nino gate}) without requiring nonlinearity.

\section{Conclusions}
\label{conclusions section}

In this paper we have investigated several designs for Bloch vector amplification gates\footnote{It would be better to call them operations or protocols because the initial conditions are always the same.} based on linear and nonlinear PTP channels, which offer benefits relative to linear CPTP channels for this application.
We do not consider microscopic models for these channels, but instead think of them as effective Markovian models for engineered strongly correlated quantum materials coupled to their environments. Thus, the models only satisfy the minimal properties of positivity and trace preservation. Our results indicate that, while non-CP dynamics is essential for fast Bloch vector amplification, NINO-type nonlinearity offers no additional computational benefit. This is because the instability underlying the gate (\ref{three-jump nino gate}) does not result from  nonlinearity but instead from a competition between gain $M$ and dissipation $\Gamma$. 

Although we have only considered  channels from class (i) and (ii), this was sufficient to achieve a significant improvement over the linear CPTP gate. In the future it would be interesting to consider amplification gates from classes (iii) and (iv) as well. Such gates might provide additional design benefits, such as robustness to noise. 
Finally, we note that the linear non-CP gate proposed here should be practical to realize, as it only requires linear operations on an open system with initial system-environment entanglement. 
It is well known that time-reversal transformations can be simulated on a quantum computer by implementing complex conjugation or reversing the sign of a simulated Hamiltonian \cite{ZhangNatComm15,LesovikSciRep19}. Similarly, fast Bloch vector amplification gates can be used to simulate a reversal of the thermodynamic arrow of time \cite{LebowitzPhysicaA93,JenningsPRE10,HanggiEnt11,220602746} for a qubit, 
and would constitute a striking demonstration physical non-CP dynamics.

\acknowledgements

This work was partly supported by the NSF under grant no. DGE-2152159.

\clearpage


\bibliography{Paper.bbl}

\end{document}